\documentclass[twocolumns, letterpaper]{aa}
\usepackage{graphicx}
\usepackage{txfonts}
\begin{document}
   \title{A variable ultra-luminous X-ray source in the colliding 
galaxy NGC\,7714}


   \author{R. Soria
          \inst{1}
          \and
          C. Motch\inst{2}}

   \offprints{R. Soria}

   \institute{Mullard Space Science Laboratory, University College London, 
	Holmbury St Mary, Surrey, RH5 6NT, UK\\
              \email{Roberto.Soria@mssl.ucl.ac.uk}
         \and
             Observatoire Astronomique, UA 1280 CNRS, 
11 rue de l'Universit\'{e}, 67000 Strasbourg, France\\
             \email{motch@astro.u-strasbg.fr}
             }

   \date{Received February 15, 2004 }

   \abstract{
We studied the colliding galaxy NGC\,7714 with two {\it XMM-Newton} 
observations, six months apart. 
The galaxy contains two bright X-ray sources: 
we show that they have different physical nature. 
The off-nuclear source is an accreting compact object,
one of the brightest ultraluminous X-ray sources (ULXs) found to date.
It showed spectral and luminosity changes between the two observations, 
from a low/soft to a high/hard state; in the high state, it reached 
$L_{\rm x} \approx 6 \times 10^{40}$ erg s$^{-1}$. 
Its lightcurve in the high state 
suggests variability on a $\approx 2$ hr timescale.
Its peculiar location, where the tidal bridge between 
NGC\,7714 and NGC\,7715 joins the outer stellar ring of NGC\,7714, 
makes it an interesting example of the connection between gas flows 
in colliding galaxies and ULX formation.
The nuclear source ($L_{\rm x} \approx 10^{41}$ erg s$^{-1}$) 
coincides with a starburst region, and is the combination 
of thin thermal plasma emission and a point-source contribution 
(with a power-law spectrum). Variability in the power-law component 
between the two observations hints at the presence 
of a single, bright point source ($L_{\rm x} \ga 3 \times 10^{40}$ 
erg s$^{-1}$): either a hidden AGN or another ULX.
   }

   \maketitle
%

\section{Introduction}

{\it XMM-Newton} and {\it Chandra} studies of colliding or merging 
gas-rich galaxies at various stages of the evolutionary sequence 
(Toomre 1977) have revealed significant contribution to the X-ray 
emission both from diffuse hot gas, associated to starburst processes, 
and from accreting point sources, generally associated to a young 
stellar population. See, for example: the Mice (Read 2003); 
the Antennae (Zezas et al.~2002; Fabbiano et al.~2003a); 
M\,82 (Griffiths et al.~2000); the Cartwheel (Gao et al.~2003).
A peculiar feature in many of these systems is the presence 
of accreting X-ray sources brighter than the Eddington limit 
for a stellar-mass black hole (BH) 
($L_{\rm Edd} \approx 10^{39}$ erg s$^{-1}$); 
they are commonly known as ultraluminous X-ray sources (ULXs).
The ages and masses of the compact objects in ULXs, 
the nature of the donor stars, 
and the geometry of emission are still unclear, and hotly debated. 
It is also unclear precisely why ULXs are preferentially 
found in interacting galaxies (Swartz et al.~2003), 
and what this can reveal about their mechanism of formation.

The interacting system Arp 284 (Arp 1966) is an exceptional example 
of a recent ($\sim 100$--$200$ Myr ago), direct impact 
(Struck \& Smith 2003). It consists of the nuclear starburst galaxy 
NGC\,7714 (classified as SB(s)b pec\footnote{NED: NASA Extragalactic Database}) 
and its fainter, currently inactive companion NGC\,7715 (Im pec).
NGC\,7714 is located at a redshift distance of $37.3$ Mpc 
(Huchra et al.~1999, for 
$H_0 = 75$ km s$^{-1}$ Mpc$^{-1}$). It has a prominent stellar ring, 
three tidal arms/tails, and is connected to NGC\,7715 by 
a gas and stellar bridge (Arp 1966). Its low inclination (viewing 
angle of $45^{\circ}$) and low foreground absorption 
($n_{\rm H} = 4.9 \times 10^{20}$ cm$^{-2}$; Dickey \& Lockman 1990) 
make it a good target for X-ray studies.

{\it ROSAT}/HRI observations of NGC\,7714 (Papaderos \& 
Fricke 1998) have revealed two strong ($L_{\rm x} > 10^{40}$ erg s$^{-1}$) 
X-ray sources, separated by $\approx 22\arcsec$. 
The brighter one coincides 
with the starburst nucleus; the fainter one is located 
approximately where the gas/stellar bridge joins a collisional 
stellar ring, but does not correlate with any bright counterpart 
at other wavelengths, nor is it located in a starburst  
region. It was suggested (Papaderos \& 
Fricke 1998) that the off-nuclear source might be 
a compact region of hot shocked gas. This could be 
due either to the collision of a fast starburst-driven outflow 
with the colder, denser gas in the galactic bridge; or, 
to the infall of high-velocity H\,{\footnotesize{I}} clouds 
along the bridge onto the outer H\,{\footnotesize{I}} disk of NGC\,7714 
(a somewhat similar situation to accretion-disk hot spots 
in X-ray binaries). 
However, the limited wavelength coverage and resolution of {\it ROSAT} 
did not allow detailed individual analyses of the two sources.

Here we present some preliminary results of our {\it XMM-Newton} 
study of the system. We argue that the off-nuclear source 
is an accreting point source (in fact, 
one of the most luminous ULXs ever detected), 
and we discuss the spectral and timing properties 
of the two sources.


   \begin{figure}
  \includegraphics[width=8.8cm]{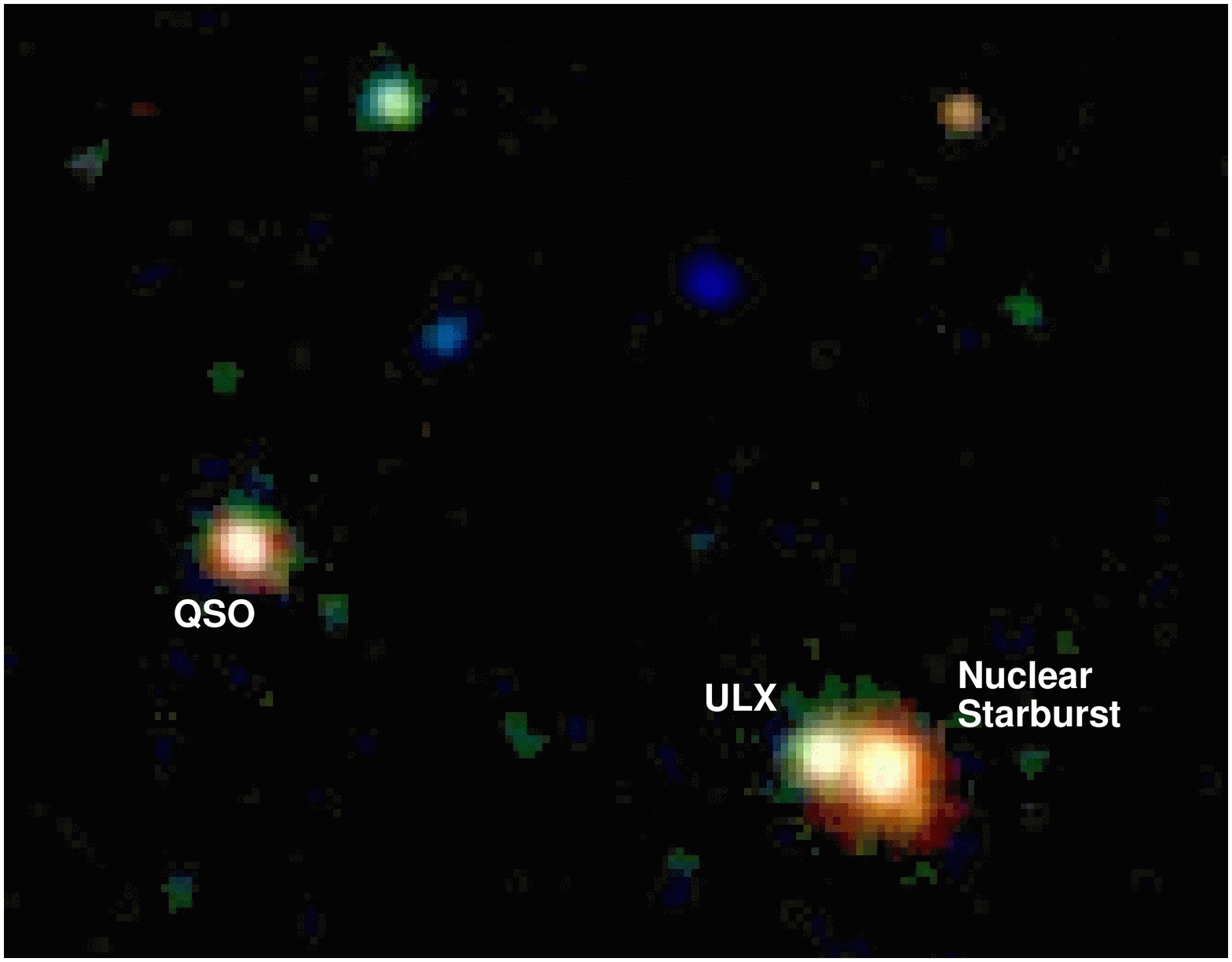}
  \includegraphics[width=8.8cm]{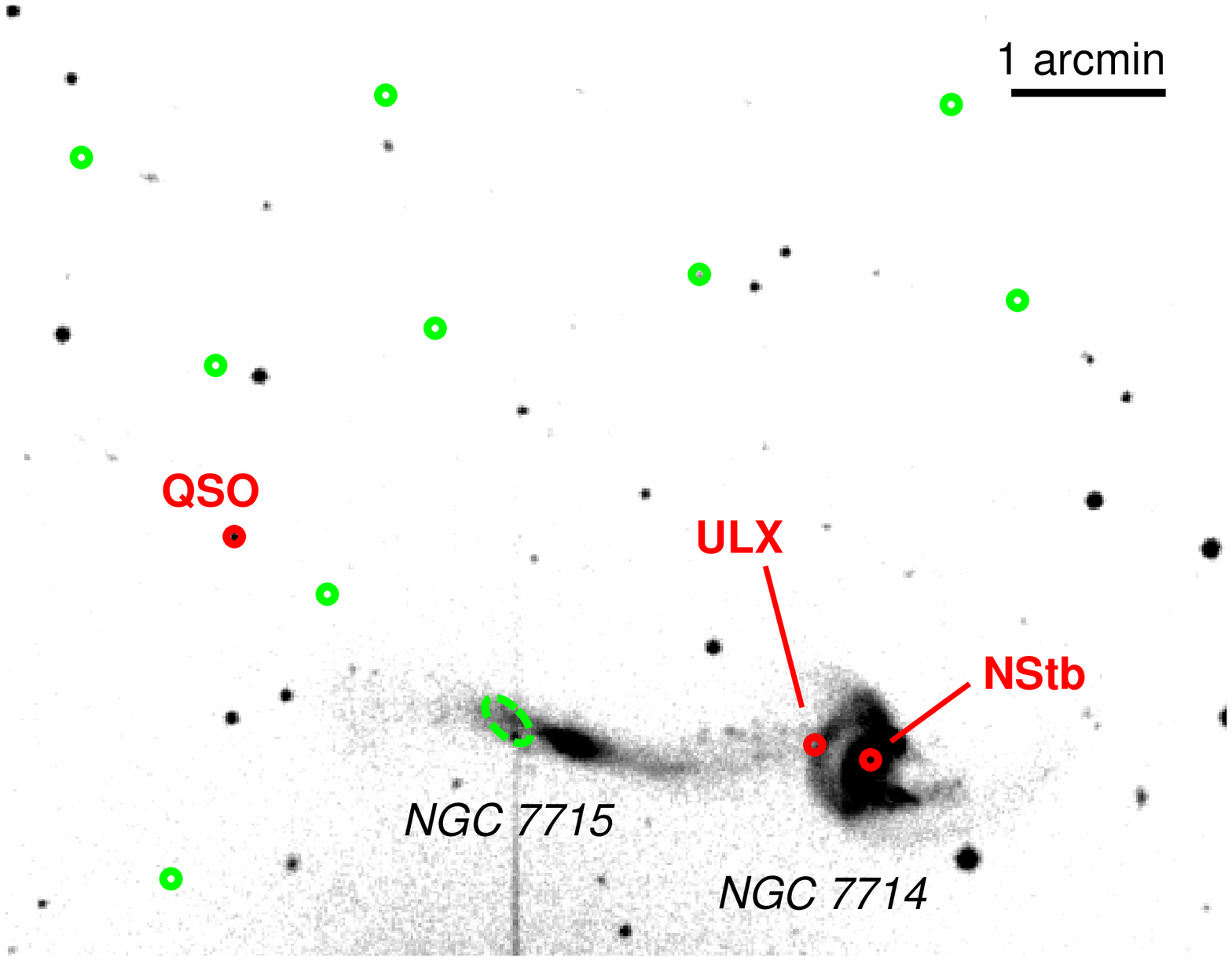}
      \caption{Top panel: true-color image of the NGC\,7714 field obtained 
by coadding the MOS and pn observations of 2002 Jun 7 and Dec 8.
The image was smoothed with a $3\times3$ pixel boxcar. Red: $0.2$--$1$ keV;
green: $1$--$2$ keV; blue: $2$--$12$ keV. Size of the image: 
$7\farcm8 \times 6\farcm2$; North is up, East is to the left.
Bottom panel: on the same scale, the positions of the three brightest 
X-ray sources in the field are overplotted (red circles) onto a Digitized 
Sky Survey $B$ image; the positions of a few other, fainter X-ray 
sources detected with {\it XMM-Newton}/EPIC are also overplotted 
(green circles and ellipse). 
              }
         \label{FigVibStab}
   \end{figure}
%

   \begin{figure}
  \includegraphics[width=9cm]{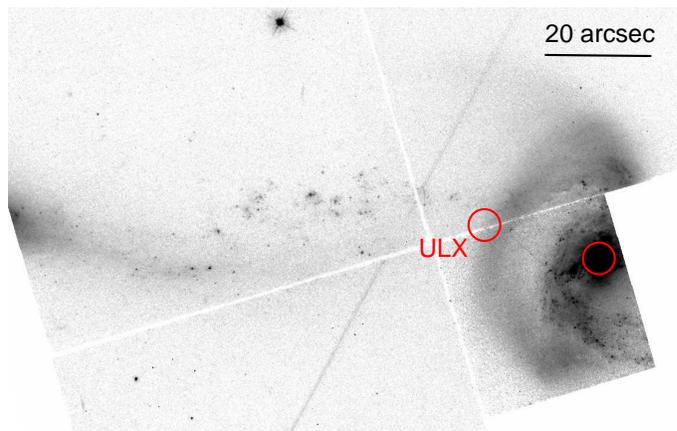}
      \caption{Position of the ULX and of the X-ray nucleus 
(red circles with radius $= 3\arcsec$) 
overplotted on an HTS/WFPC2 image in the $V$ band (f555w filter).
The ULX is located where the tidal bridge meets the collisional 
stellar ring. North is up, East is to the left.
              }
         \label{FigVibStab}
   \end{figure}
%

%

\section{Observations and data analysis}

NGC\,7714 was observed with all instruments on-board 
{\it XMM-Newton} on 2002 June 7 and 2002 December 8 (Table 1);
we used the medium filter, full-frame mode for the EPIC detectors.
We processed the Observation Data Files with standard tasks 
in Version 5.4 of the Science Analysis System (SAS). 
After inspecting the background fluxes, we rejected 
the initial $\approx 1$ ks of the first exposure, 
which was affected by a background flare: we retained  
a live-time good-time-interval of 15.9 ks (for the MOSs) 
and 14.3 ks (for the pn). From the second observation,
we used a live-time good-time-interval of 15.8 ks (MOSs) and 12.9 ks (pn).
We filtered the event files, selecting only the best-calibrated 
events (pattern $\leq 12$ for the MOSs, pattern $\leq 4$ for pn), and 
rejecting flagged events. We checked and corrected the astrometry 
of the various images so that the position of the nuclear source 
coincided with the NED position (Clements 1983): 
 {R.A.~(2000) $=$ 23$^h$\,36$^m$\,$14^s.1$}, 
{Dec.~(2000) $=$ $+$02$^{\circ}$\,09\arcmin\,18\farcs6}.

Firstly, we extracted source and background spectra 
for the nucleus and the bright off-nuclear source 
in NGC\,7714, for each detector in each of the two exposures. 
The off-nuclear source is located at 
{R.A.~(2000) $=$ 23$^h$\,36$^m$\,$15^s.6$}, 
{Dec.~(2000) $=$ $+$02$^{\circ}$\,09\arcmin\,23\farcs5}, 
ie, $\approx 22\arcsec$~ east of the nucleus (Figs.~1,2).
We estimate an error $\la \pm 3\arcsec$, mostly due 
to the astrometry uncertainties; however, the error 
in the relative positions of the nuclear 
and off-nuclear sources is only $\approx 1\arcsec$.
To reduce the contamination from the bright nearby 
nucleus, we used as source extraction region a circle 
of radius $11\arcsec$. We used as background region 
the union of three circles of $11\arcsec$ radius, centred 
at a distance of $22\arcsec$~from the nucleus, 
on the opposite side of the source, so that it would 
contain a similar contribution from the nuclear source.
For the nucleus, we used an extraction circle of radius $15\arcsec$. 
The background region was chosen as the union of two circles 
of radius $15\arcsec$~centred at $22\arcsec$~from the nearby 
off-nuclear source. We tested the effect of different choices of 
background regions, until we were satisfied that the reciprocal 
contaminations of the two sources were properly subtracted.

We built response functions for each source with 
the SAS task {\tt rmfgen}, and auxiliary response 
functions with {\tt arfgen}. Given the small 
extraction radii used for the two overlapping sources, 
we corrected for the flux that is outside the extraction 
regions by setting ``extendedsource=false'', ``modelee=true'' 
in the {\tt arfgen} parameters.
We then fitted the background-subtracted 
spectra with standard models in {\footnotesize XSPEC v.11.3} (Arnaud 1996); 
owing to the uncertainties in the EPIC responses at low energies, 
we used only the $0.3$--$12$ keV range.
Firstly, we analysed the individual pn and MOS spectra from each observation. 
After ascertaining that they were consistent with each other, 
we coadded them with the method described in Page et al.~(2003), 
in order to increase the signal-to-noise ratio. 
Thus, we obtained one combined EPIC spectrum of the nucleus and one 
of the off-nuclear source from each epoch.

   \begin{table}
      \caption[]{{\it XMM-Newton}/EPIC observation log for NGC\,7714.}
         \label{table1}
\begin{centering}
         \begin{tabular}{lrrrr}
            \hline
            \hline
            \noalign{\smallskip}
            Date & Obs ID & \hspace{-0.2cm}Start time & \hspace{-0.2cm}Stop time & \hspace{-0.4cm}GTI\\
		&&&&(ks)\\[3pt]
            \noalign{\smallskip}
	    \hline
            \hline
            \noalign{\smallskip}
            \noalign{\smallskip}
		2002 Jun 07 & 0112521301 & pn: 10:36:39 & 15:16:39 & 15.8 \\
			& & MOS: 10:03:23 & 15:21:23 & 16.1\\[5pt]
            	2002 Dec 08 & 0112522601 & pn: 12:40:16 & 16:38:19 & 14.3\\
			&	&	  MOS: 12:18:15 & 16:43:15 & 15.9\\
            \noalign{\smallskip}
	    \hline
         \end{tabular}
\end{centering}
   \end{table}

%

%

   \begin{figure}
  \includegraphics[angle=-90,width=8.5cm]{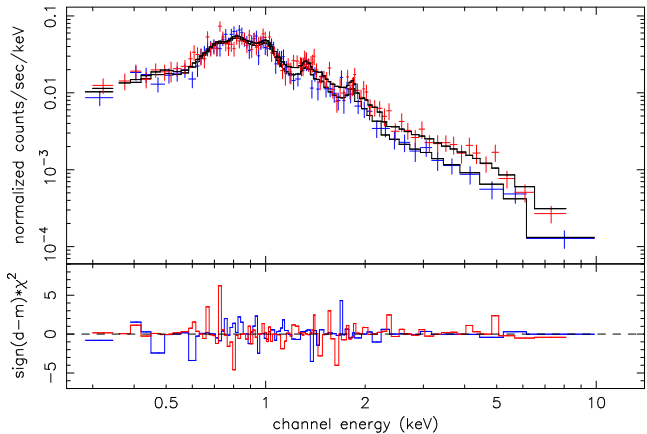}
      \caption{Coadded EPIC spectra of the nucleus on 2002 Jun 7 (blue) 
and Dec 8 (red), fitted with a constant two-temperature vmekal model 
plus a varying power-law component.
              }
         \label{FigVibStab}
   \end{figure}
%

   \begin{figure}
  \includegraphics[angle=-90,width=8.5cm]{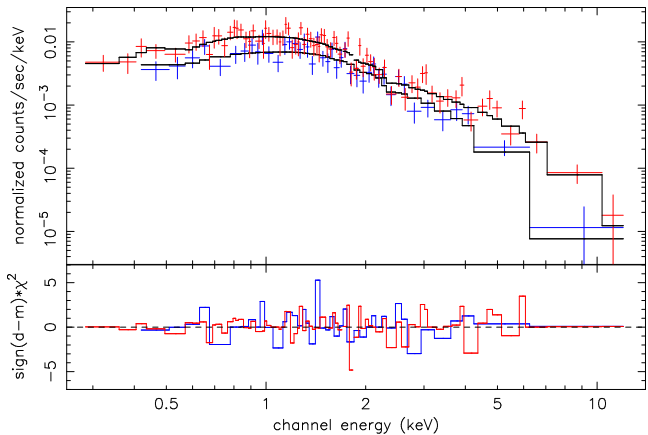}
      \caption{Coadded EPIC spectra of the ULX on 2002 Jun 7 and Dec 8, 
fitted with a disk-blackbody and a power-law, respectively.
              }
         \label{FigVibStab}
   \end{figure}
%

\section{Main results}

\subsection{Starburst Nucleus}

The nucleus of NGC\,7714 is characterised by an active 
starburst that started $\sim$ a few $\times 10^8$ yr ago, 
with an average star-formation rate $\approx 1 M_{\odot}$ yr$^{-1}$; 
the rate has probably been higher by a factor of a few 
during the most recent burst which started $\approx 5$ Myr ago 
(Lan\c{c}on et al.~2001). 
Its optical, UV and IR properties are similar to those 
of the ``prototypical'' starburst galaxy M\,82.
In analogy with other starburst galaxies, we expect the nuclear 
X-ray emission to be a combination of diffuse hot gas, 
high-mass X-ray binaries and supernova 
remnants---including perhaps some contribution 
from the Type Ib/c SN\,1999dn, located $\approx 13\arcsec$ 
south-east of the nucleus (Qiu et al.~1999; Deng et al.~2000). 

The limited spatial resolution 
of {\it XMM-Newton} does not allow us to resolve individual 
sources in the nuclear region. However, we checked whether 
the observed profile is consistent with the point-spread-function 
of a point source, using standard tasks in {\footnotesize IRAF}. 
We compared the full-width half-maximum (FWHM)
of the radial profile for the nucleus, the off-nuclear source 
and a nearby quasar. Using EPIC/MOS images from the two observations, 
we obtain that the last two sources have FWHM $\approx 6\arcsec$, 
while the nucleus has FWHM $\approx 8\arcsec$. This suggests 
that the X-ray emission from the nuclear region is indeed extended,  
in agreement with the {\it ROSAT}/HRI results 
(Papaderos \& Fricke 1998).
No difference in the radial extent of the three sources 
is found in the EPIC/pn images; however, this is due  
to the lower spatial resolution and larger pixel size of the pn.
We also examined the radial profiles separately for ``soft'' 
($0.2$--$1.5$ keV) and ``hard'' ($1.5$--$12$ keV) MOS images.
We obtain that, in the hard band, the FWHM is the same 
for all three objects, consistent with point sources.
The larger radial extent of the nuclear source is found  
only in the soft band, suggesting a different origin 
for the two components (see Section 4.3). Further investigation 
of the MOS data suggests that the larger radial profile 
of the nuclear source in the soft band is due 
to faint emission a few arcsec south-west 
of the central position, in excess of what would be expected 
from a point-source point-spread-function. 
No excess emission is detected east of the central position, 
suggesting that SN\,1999dn does not contribute significantly.

We fitted the nuclear X-ray spectrum with an absorbed 
two-temperature variable-abundance thermal plasma model 
({\tt vmekal}), plus an absorbed power-law 
component, which accounts for the contribution from 
accreting compact objects.
We find that we cannot fit the EPIC spectra from 2002 June and December 
simultaneously with the same set of parameters. 
The two spectra are identical at energies $\la 1$ keV, 
but the flux at higher energies was higher in December.
Therefore, we fixed the two thermal-plasma components 
(which we do not expect to vary on a 6-months' timescale)
but allowed the power-law contribution to vary between 
the two observations. This results in a good fit ($\chi^2_\nu = 0.77$, 
see Table 2 and Fig.~3). We do not obtain statistically-significant 
improvements to the fit by allowing different normalisations 
of the thermal plasma components in the two observations. 
The total emitted luminosity in the $0.3$--$12$ keV band 
is $\approx 8 \times 10^{40}$ erg s$^{-1}$ in June and 
$\approx 1.1 \times 10^{41}$ erg s$^{-1}$ in December.
The fitted temperatures of the optically-thin thermal components are 
$kT_1 = 0.43^{+0.10}_{-0.16}$ keV and $kT_2 =0.98^{+3.40}_{-0.15}$ keV. 
Thermal plasma emission contributes for $\approx 30$--$40\%$ 
of the emitted luminosity in the $0.3$--$12$ keV band 
and $\approx 75$--$80\%$ in $0.3$--$1$ keV band.
The power-law component has a slope $\Gamma \approx 1.9$, 
consistent both with an AGN spectrum and with typical 
X-ray binary spectra. 
The emitted luminosity in the $2$--$10$ keV band 
is $\approx 2.2 \times 10^{40}$ erg s$^{-1}$ 
and $\approx 3.8 \times 10^{40}$ erg s$^{-1}$ in the two epochs: 
this is consistent with a current star-formation rate 
of a few $M_{\odot}$ yr$^{-1}$ (Gilfanov et al.~2004).


   \begin{table}
      \caption[]{Best-fit parameters for the combined EPIC spectra 
of the nuclear source on 2002 Jun 7 and Dec 8. We assumed a Galactic column 
density $n_{\rm H,Gal} = 4.9 \times 10^{20}$ cm$^{-2}$. We fitted 
the two spectra simultaneously, assuming that the thermal plasma 
component did not vary between the two epochs, but leaving the power-law 
component free to vary. The quoted errors 
are the 90\% confidence limit.}
         \label{table1}
\begin{centering}
         \begin{tabular}{lrr}
            \hline
            \hline
            \noalign{\smallskip}
            Parameter      &  Value in Jun 02&  Value in Dec 02\\[3pt]
            \noalign{\smallskip}
	    \hline
            \hline
            \noalign{\smallskip}
            \noalign{\smallskip}
	\multicolumn{3}{c}{model: wabs$_{\rm Gal}~\times$ 
		(wabs$_{\rm vm}~\times$ (vmekal$_1$~$+$~vmekal$_2$)~
		$+$~wabs$_{\rm po}~\times$ po)}\\
            \noalign{\smallskip}
            \hline
            \noalign{\smallskip}
            \noalign{\smallskip}
$n_{\rm H,vm}~(10^{21}$~cm$^{-2}$)  & \multicolumn{2}{r}{$0.8^{+4.3}_{-0.3}$\hspace*{1.4cm}} \\[3pt] 
$n_{\rm H,po}~(10^{21}$~cm$^{-2}$)  & \multicolumn{2}{r}{$1.0^{+1.0}_{-0.6}$\hspace*{1.4cm}}\\[3pt] 
$kT_{\rm vm1}$~(keV)    & \multicolumn{2}{r}{$0.43^{+0.10}_{-0.16}$\hspace*{1.4cm}} \\[3pt]
$kT_{\rm vm2}$~(keV)    & \multicolumn{2}{r}{$0.98^{+3.40}_{-0.15}$\hspace*{1.4cm}}\\[3pt]
$K_{\rm vm1}~(10^{-5})$   & \multicolumn{2}{r}{$3.6^{+9.8}_{-1.5}$\hspace*{1.4cm}} \\[3pt] 
$K_{\rm vm2}~(10^{-5})$   & \multicolumn{2}{r}{$2.0^{+1.8}_{-1.4}$\hspace*{1.4cm}}  \\[3pt]
$\Gamma$ & $2.00^{+0.22}_{-0.21}$ & $1.85^{+0.19}_{-0.19}$  \\[3pt]
$K_{\rm po}~(10^{-5})$ & $4.3^{+1.4}_{-1.0}$ & $6.0^{+1.6}_{-1.4}$   \\[3pt]
C     &  \multicolumn{2}{r}{$(0.8^{+0.9}_{-0.6})^{\mathrm{~a}}$\hspace*{1.2cm}} \\[3pt]
N  &\multicolumn{2}{r}{$(0.8^{+0.9}_{-0.6})^{\mathrm{~a}}$\hspace*{1.2cm}} \\[3pt]
O  &\multicolumn{2}{r}{$(0.8^{+0.9}_{-0.6})^{\mathrm{~a}}$\hspace*{1.2cm}} \\[3pt] 
Ne  & \multicolumn{2}{r}{$1.2^{+1.2}_{-1.2}$\hspace*{1.4cm}} \\[3pt]
Na  & \multicolumn{2}{r}{$(1.0)$\hspace*{1.4cm}}\\[3pt]
Mg  & \multicolumn{2}{r}{$2.9^{+2.3}_{-1.5}$\hspace*{1.4cm}} \\[3pt]
Al  & \multicolumn{2}{r}{$(1.0)$\hspace*{1.4cm}} \\[3pt]
Si  & \multicolumn{2}{r}{$3.6^{+5.4}_{-1.7}$\hspace*{1.4cm}}   \\[3pt]
S  & \multicolumn{2}{r}{$(1.0)$\hspace*{1.4cm}} \\[3pt]
Ar & \multicolumn{2}{r}{$(1.0)$\hspace*{1.4cm}}  \\[3pt]
Ca  & \multicolumn{2}{r}{$(1.0)$\hspace*{1.4cm}} \\[3pt]
Fe  & \multicolumn{2}{r}{$(1.0)$\hspace*{1.4cm}}\\[3pt]
Ni  & \multicolumn{2}{r}{$(1.0)$\hspace*{1.4cm}} \\[3pt] 
            \noalign{\smallskip}
            \hline
            \noalign{\smallskip}
		$\chi^2_\nu$ & \multicolumn{2}{r}{$0.77~(122.4/160)$\hspace*{0.4cm}} \\[3pt]
            	$f_{0.3{\rm -}12}~(10^{-13}~{\rm erg~cm}^{-2}~{\rm s}^{-1})$ 
			& $2.8^{+0.3}_{-0.8}$ & $4.0^{+0.2}_{-0.8}$\\[3pt]
            	$f_{0.3{\rm -}1}~(10^{-13}~{\rm erg~cm}^{-2}~{\rm s}^{-1})$ 
			& $0.8^{+0.1}_{-0.4}$ & $0.9^{+0.1}_{-0.4}$\\[3pt]
            	$L_{0.3{\rm -}12}~(10^{40}~{\rm erg~s}^{-1})$ & $7.8$  &$10.5$  \\
            	$L_{0.3{\rm -}12,{\rm vm}}~(10^{40}~{\rm erg~s}^{-1})$ & \multicolumn{2}{r}{$2.9$\hspace*{1.4cm}} \\
            	$L_{0.3{\rm -}12,{\rm po}}~(10^{40}~{\rm erg~s}^{-1})$ & $4.9$  &$7.6$  \\
            	$L_{0.3{\rm -}1}~(10^{40}~{\rm erg~s}^{-1})$ & $3.5$  & $3.9$ \\
            	$L_{2{\rm -}10}~(10^{40}~{\rm erg~s}^{-1})$ & $2.2$  & $3.8$ \\
            \noalign{\smallskip}
	    \hline
         \end{tabular}
\begin{list}{}{}
\item[$^{\mathrm{a}}$] We imposed the same abundance for these three elements. 
\end{list}
\end{centering}
   \end{table}

As an aside, we also searched for X-ray emission from NGC\,7715, 
the tidally-interacting companion to NGC\,7714. Current star-formation 
from this galaxy is known to be negligible, probably 
because of a relative lack of gas (Struck \& Smith 2003). 
We did not significantly detect 
any source from each observation; however, when we combine 
both the June and December datasets, we detect a faint, possibly extended 
source at $\approx 3$-$\sigma$ level in the EPIC image (Fig.~1). 
Its location ({R.A.~(2000) $=$ 23$^h$\,36$^m$\,$23^s.6$}, 
{Dec.~(2000) $=$ $+$02$^{\circ}$\,09\arcmin\,32\arcsec}) puts it 
in the eastern gas/stellar countertail of NGC\,7715 rather than 
in its nucleus. From the detected count rate, we estimate 
an average luminosity $\approx 10^{39}$ erg s$^{-1}$.

\subsection{Off-nuclear ULX}

We then examined the coadded EPIC spectra of the off-nuclear source 
from the two observations. 
It is immediately clear that this source 
is fundamentally different from the starburst nucleus: its X-ray spectrum 
is featureless, with no significant evidence of optically-thin plasma emission. 
It is also clear that the source varied significantly between 
the two epochs.

The source cannot be well fitted with a single spectral model 
in both observations. Limiting our choice to simple models, 
we find that a multicolor disk-blackbody with $kT_{\rm in} = 1.0\pm0.1$ keV 
provides the best fit to the 2002 June spectrum ($\chi^2_\nu = 1.11$; see 
Table 3 and Fig.~4). 
A simple power-law fit gives $\Gamma = 2.5\pm0.2$ ($\chi^2_\nu = 1.20$), 
but is slightly improved by the addition of a blackbody component
at $kT =0.66^{+0.35}_{-0.24}$ keV ($\chi^2_\nu = 1.15$).
On the contrary, a simple power-law model with $\Gamma = 2.1^{+0.2}_{-0.1}$ 
is an excellent fit to the 2002 December spectrum ($\chi^2_\nu = 0.86$); 
adding a thermal component does not improve the fit.
A multicolor disk-blackbody model is clearly rejected ($\chi^2_\nu = 1.29$) 
for the second observation. 
We also tried fitting the two spectra with the Comptonisation 
model {\tt bmc} (Shrader \& Titarchuk 1999) (Table 3), confirming 
that the source was softer in June.

Although any estimates of the emitted luminosity 
are very model-dependent, it is clear that the source 
has become brighter by a factor of $\sim 2$ 
between the two epochs, reaching a luminosity $\approx 6 \times 10^{40}$ 
erg s$^{-1}$ in 2002 December.\footnote{
The fact that both the ULX and the nuclear source 
showed an increase in luminosity from June to December 
prompted us to double-check that this effect is not due 
to systematic errors. The change in flux is at least 
an order of magnitude greater than any possible 
reciprocal contamination between the two sources.
Moreover, we verified that the luminosity and spectral 
variations are consistently seen in the individual 
pn and MOS spectra; this rules out a systematic error 
in our coadding of the datasets.
In the nucleus, we can clearly attribute the change 
to the power-law component only, while the thermal plasma 
emission is unchanged: this also confirms that it is not 
an instrumental artifact. As a further check, 
we extracted and analysed the spectra of the nearby QSO 
$\left[{\rm HB}89\right]\,2333+019$ (Fig.~1)
for the two epochs. We find that the best-fit parameters 
(power-law with $\Gamma = 1.65^{+0.11}_{-0.11}$, 
and $\Gamma = 1.79^{+0.13}_{-0.10}$ respectively) 
are consistent with an unchanged spectrum. 
In both observations, the detected QSO flux in the $0.3$--$12$ keV band was 
$f_{\rm x} = 2.7 \times 10^{-13}$ erg~cm$^{-2}$ s$^{-1}$. 
We conclude that the variations seen 
in the ULX and the nucleus between 2002 June and December are real.}
Such variability on a 6-month timescale 
suggests that the size of the emitting region 
is $\la 0.2$ pc.


In addition to the long-term variability of the ULX flux, 
we looked for short-term variations during the two observations. 
We find that the background-subtracted $0.2$--$12$ keV count rate 
is consistent with a constant level in the low state. However, variability 
is detected in the high state (Fig.~5), significant at the $90\%$ level 
according to the Kolmogorov-Smirnov test.
Fitting the data with a {\it sin} curve gives 
a period of $6920^{+1300}_{-900}$ s, and a semi-amplitude 
of $\approx 18\%$. However, longer observations are needed 
to ascertain whether the lightcurve has a real periodicity.
No significant variability is detected from the nucleus within 
each of the two observations.

   \begin{table}
      \caption[]{Best-fit parameters for the combined EPIC spectra 
of the ULX on 2002 Jun 7 and Dec 8. The quoted errors 
are the 90\% confidence limit. We assumed a Galactic column 
density $n_{\rm H,Gal} = 4.9 \times 10^{20}$ cm$^{-2}$.}
         \label{table1}
\begin{centering}
         \begin{tabular}{lrr}
            \hline
            \hline
            \noalign{\smallskip}
            Parameter      &  Value in Jun 02 & Value in Dec 02 \\[3pt]
            \noalign{\smallskip}
	    \hline
            \hline
            \noalign{\smallskip}
            \noalign{\smallskip}
	\multicolumn{3}{c}{model: wabs$_{\rm Gal}~\times$ 
		wabs $\times$ (bb $+$ po)}\\
            \noalign{\smallskip}
            \hline
            \noalign{\smallskip}
            \noalign{\smallskip}
            	$n_{\rm H}~(10^{21}~{\rm cm}^{-2})$ & $2.2^{+2.1}_{-1.1}$ & $1.5^{+0.4}_{-0.4}$     \\[3pt]
		$kT_{\rm bb}~({\rm keV})$ & $0.66^{+0.35}_{-0.24}$ & -\\[3pt]
		$K_{\rm bb}~(10^{-7})$ & $4.4^{+4.2}_{-3.5}$ & -\\[3pt]
		$\Gamma$  & $2.6^{+0.4}_{-0.7}$ & $2.1^{+0.2}_{-0.1}$\\[3pt]
		$K_{\rm po}~(10^{-5})$ & $3.8^{+2.3}_{-1.3}$ & $6.1^{+1.1}_{-0.9}$\\
            \noalign{\smallskip}
            \hline
            \noalign{\smallskip}
		$\chi^2_\nu$ & $1.15\,(43.7/38)$ & $0.86\,(64.8/75)$\\[3pt]
            	$f_{0.3{\rm -}12}~(10^{-13}~{\rm erg~cm}^{-2}~{\rm s}^{-1})$ 
			& $1.1^{+0.2}_{-0.5}$ & $2.3^{+0.2}_{-0.2}$\\[3pt]
            	$L_{0.3{\rm -}12}~(10^{40}~{\rm erg~s}^{-1})$ & $4.3$ & $6.6$     \\
            \noalign{\smallskip}
	    \hline
            \noalign{\smallskip}
            \noalign{\smallskip}
	\multicolumn{3}{c}{model: wabs$_{\rm Gal}~\times$ 
		wabs $\times$ diskbb}\\
            \noalign{\smallskip}
            \hline
            \noalign{\smallskip}
            \noalign{\smallskip}
		$n_{\rm H}~(10^{21}~{\rm cm}^{-2})$ & $0.07^{+0.49}_{-0.07}$ & $<0.12$ \\[3pt]
		$kT_{\rm in}~({\rm keV})$ & $0.97^{+0.12}_{-0.13}$ & $1.07^{+0.13}_{-0.10}$\\[3pt]
		$K_{\rm dbb}~(10^{-3})$ & $6.2^{+4.9}_{-2.1}$ & $7.4^{+2.6}_{-2.5}$\\
            \noalign{\smallskip}
            \hline
            \noalign{\smallskip}
		$\chi^2_\nu$ & $1.11\,(44.4/40)$ & $1.29\,(96.7/75)$\\[3pt]
            	$f_{0.3{\rm -}12}~(10^{-13}~{\rm erg~cm}^{-2}~{\rm s}^{-1})$ 
			& $1.0^{+0.1}_{-0.2}$ & $1.8^{+\ast}_{-\ast}$\\[3pt]
            	$L_{0.3{\rm -}12}~(10^{40}~{\rm erg~s}^{-1})$ & $2.3$ & $3.8$     \\
            \noalign{\smallskip}
	    \hline
            \noalign{\smallskip}
            \noalign{\smallskip}
	\multicolumn{3}{c}{model: wabs$_{\rm Gal}~\times$ 
		wabs $\times$ bmc}\\
            \noalign{\smallskip}
            \hline
            \noalign{\smallskip}
            \noalign{\smallskip}
		$n_{\rm H}~(10^{21}$~cm$^{-2})$ &$<0.37$ & $0.34^{+2.9}_{-0.34}$\\[3pt] 
		$kT_{\rm bb}$~(keV)   & $0.34^{+0.04}_{-0.04}$ & $0.22^{+0.11}_{-0.22}$\\[3pt]
		$\Gamma$         & $2.9^{+0.4}_{-0.3}$ & $1.9^{+0.3}_{-0.4}$\\[3pt]		
		log(A)         & $>0.10$ & $>0.21$\\[3pt]
		$K_{\rm bmc}~(10^{-7})$ & $9.4^{+0.7}_{-0.8}$
			& $14.5^{+5.0}_{-3.4}$\\
            \noalign{\smallskip}
            \hline
            \noalign{\smallskip}
		$\chi^2_\nu$ & $1.16\,(44.1/38)$ & $0.87\,(63.6/73)$\\[3pt]
            	$f_{0.3{\rm -}12}~(10^{-13}~{\rm erg~cm}^{-2}~{\rm s}^{-1})$ 
			& $1.1^{+0.1}_{-\ast}$ & $2.4^{+0.1}_{-0.6}$\\[3pt]
            	$L_{0.3{\rm -}12}~(10^{40}~{\rm erg~s}^{-1})$ & $2.4$ & $5.1$     \\
            \noalign{\smallskip}
	    \hline
         \end{tabular}
\end{centering}
   \end{table}

   \begin{figure}
  \includegraphics[angle=-90,width=8.5cm]{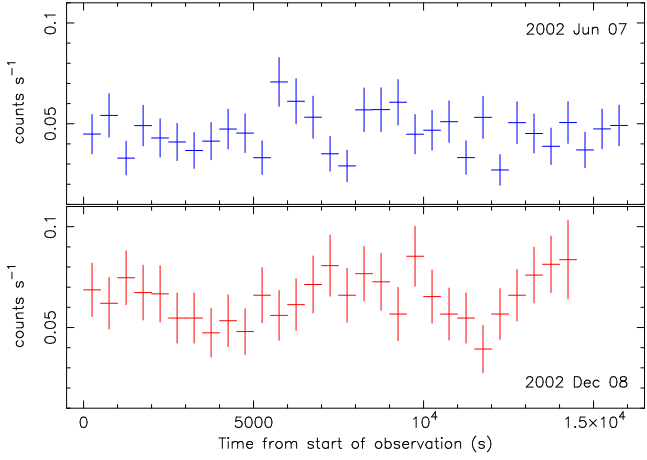}
      \caption{EPIC lightcurve for the ULX on 2002 Jun 7 and Dec 8, 
binned to 500 s. 
The background-subtracted count rate is the combined MOS plus pn count 
rate within an $11\arcsec$ extraction region; 
the encircled energy within that radius 
is $\approx 59\%$ for MOS1, and $\approx 54\%$ for MOS2 and pn.             }
         \label{lc}
   \end{figure}
%

\section{Discussion}

\subsection{Intermediate-mass BH or beamed emission?}

It was suggested (Papaderos \& Fricke 1998) that the off-nuclear 
X-ray source could be emission from thermal plasma in a hot spot, 
due either to the collision of a fast outflow with the denser gas 
in the galactic bridge; or, to the infall of high-velocity clouds 
along the bridge onto the outer H\,{\footnotesize{I}} disk of NGC\,7714. 
However, both those hypotheses are now ruled out 
by our {\it XMM-Newton} study.
From the X-ray spectral properties in the two observations, 
and from the variability between the two epochs, we conclude 
that the off-nuclear source is instead an accreting compact object. 
As such, with an extrapolated bolometric luminosity $\approx 1.5 \times 10^{41}$ 
erg s$^{-1}$ in the high state, it is one of the brightest ULXs ever found.
If the emission is isotropic, satisfying the Eddington limit 
would require a mass of $\approx$ a few $\times 10^2 M_{\odot}$ 
for the accreting BH.

The observed transition between a low/soft and a high/hard 
spectral state is opposite to what is generally observed in Galactic 
stellar-mass BH candidates, but similar to what was found 
for the ULX Holmberg II X-1 (Dewangan et al.~2004) 
and for some ULXs in the Antennae (Fabbiano et al.~2003b). 
In the high state, the X-ray spectrum  
can be modelled by a pure power law with photon index $\Gamma \approx 2$; 
the conventional explanation 
for this component is inverse-Compton scattering of soft photons 
in a hot corona (Sunyaev \& Titarchuk 1980).

A thermal component (blackbody or disk-blackbody) 
is instead significantly detected in the low/soft state:  
it contributes for $\approx (0.5$--$2.3) \times 10^{40}$ 
erg s$^{-1}$ depending on the choice of spectral model; 
the short duration of our {\it XMM-Newton} 
observations does not allow us to put stronger constraints 
on the thermal emission. When the low-state spectrum is fitted 
with a disk-blackbody model, the color-temperature 
parameter $kT_{\rm in} \approx 1$ keV (Table 3). If this parameter 
is identified with the effective temperature 
at the inner boundary of a Shakura-Sunyaev disk 
(Shakura \& Sunyaev 1973), its high value is inconsistent 
with the BH mass inferred from the X-ray luminosity.
High disk temperatures inconsistent with the standard disk model 
have been found in many other ULXs (Makishima et al.~2000).
They have been explained (Mineshige et al.~2000; 
Watarai et al.~2001) with a different 
structure of the accretion flow at high accretion rates 
(``slim-disk''). Using the slim-disk model, our  
fitted value of $T_{\rm in}$ corresponds to a BH mass 
$\approx 200 M_{\odot}$ (in the Schwarzschild case), 
with a mass accretion rate $\approx 3 \times 10^{-5} M_{\odot}$ yr$^{-1}$.
Another suggestion put forward to explain the high color 
temperatures in ULXs is a correction 
to the disk-blackbody spectrum via a hardening factor 
$T_{\rm in}/T_{\rm eff} \approx 2.6$ 
(Shrader \& Titarchuk 2003). Finally, a standard accretion disk 
around a fast-rotating Kerr BH would extend closer to the event horizon 
and therefore would also have 
a higher inner temperature than predicted in the non-rotating 
(Schwarzschild) case (Zhang et al. 1997).
However, this effect would be significant only at high inclination 
angles (edge-on systems) and is unlikely to be 
the correct explanation for the high-temperature 
ULX disk spectra (Ebisawa et al.~2003).

An alternative to the intermediate-mass BH scenario 
is the possibility that we are looking into a very narrowly collimated beam 
of emission (beaming factor $\ga 50$) from a stellar-mass X-ray binary 
(BH mass $\la 20 M_{\odot}$).
Although the latter scenario cannot be ruled out a priori, 
it would be a very peculiar coincidence to find that the only strongly 
collimated source aligned---by chance---with our line of sight 
is located precisely at the junction between the collisional 
stellar ring and the tidal bridge. 
A more general, quantitative argument against simple geometrical 
beaming is discussed in Davis \& Mushotzky (2004), 
based on the consideration that the scattering region 
(thick disk or torus) responsible for the beaming can never be 
a perfect mirror; a large fraction of the radiation would 
be absorbed and re-radiated isotropically in the optical/IR. 
Hence, one would expect to find many bright (absolute 
magnitudes up to $\approx -11$) optical/IR point sources 
for every ULX. Such bright sources have never been seen 
in nearby galaxies.

Alternatively, the observed ultraluminous flux might be due 
to relativistic beaming (Urry \& Shafer 1984; 
Georganopoulos et al.~2002; K\"{o}rding et al.~2002). In this scenario, 
the power-law component in the X-ray spectrum 
is the result of inverse-Compton scattering of soft photons 
by the electrons in a jet, moving with velocity $\beta$ and 
a bulk motion factor $\Gamma = (1-\beta^2)^{-1/2}$.
Relativistic Doppler boosting implies that 
the observed luminosity in a given band is enhanced 
by a factor $\left[\Gamma \, (1-\beta\cos\theta)\right]^{-4}$.
Assuming that the accreting source in NGC\,7714 is a stellar-mass BH, 
and that the mechanical luminosity in the jet is $\la 0.1 L_{\rm Edd}$ 
(eg, Fender~2001), Lorentz factors $\Gamma \ga 3$ 
and beaming angles $\la 5^{\circ}$ are required 
to explain the observed luminosities in the high/hard state.
However, this scenario would not explain the soft thermal component
seen in the low state, which is unlikely to be beamed.

In conclusion, the location, luminosity and spectral properties 
of the ULX favour a massive ($M> 100 M_{\odot}$) 
accreting object. Further constraints can come 
from time-variability studies. 
For example, a break in the power-density spectrum (PDS) 
at a frequency $\approx (M/M_{\odot}$) Hz has been detected 
in some AGN and Galactic BH binaries (for recent 
examples see Markowitz et al.~2003; 
Uttley et al.~2002; Czerny et al.~2001), 
and from a ULX in NGC\,4559 (Cropper et al.~2004).
A preliminary PDS analysis of the ULX in NGC\,7714 in its high state 
suggests features at $\sim 1.5 \times 10^{-4}$ Hz (also apparent 
from the lightcurve in Fig.~5) and $\sim 2 \times 10^{-4}$ Hz; 
hence, longer observations could be interesting.

\subsection{ULX formation in colliding galaxies}

One of the few statistically significant properties 
of the ULX population in nearby galaxies is that these 
sources are more frequently found in merging or tidally 
interacting galaxies (Swartz et al.~2003).
It is not yet clear why. A simple suggestion is that the 
ULX formation rate is proportional 
to the star-formation rate (eg, Gilfanov et al.~2003),
which is known to be higher in interacting galaxies. 
More specifically, ULX formation could be related 
to the clustered star-formation rate.

The ULX in NGC\,7714 is located on the outer stellar ring, 
at the junction with the tidal bridge. 
ULXs are often found in collisional rings and tails 
of tidally disrupted systems, for example 
in the outer stellar ring of the Cartwheel galaxy (Gao et al.~2003).
However, the two situations are fundamentally different.
In the Cartwheel, the outer ring is defined by an expanding 
wave of star formation (Higdon 1995) associated to a strong density wave 
in the gas, triggered by the collisional perturbation;
hence, ULXs in the Cartwheel ring are clearly associated 
with active star formation and young stellar clusters.
Here, instead, the off-centre collision between the two galaxies produced 
a weaker dynamical perturbation (Struck \& Smith 2003), 
and the collisional outer ring is only an expanding density wave 
of old stars (Bushouse \& Werner 1990), without 
ongoing star formation: no H$\alpha$ emission 
is observed along the ring (Gonz\'{a}lez-Delgado et al.~1995; 
Smith et al.~1997). Radio 21-cm VLA observations 
also do not show any H{\footnotesize I} 
counterpart to the ring (Smith et al.~1997).

Moderately active star formation and H{\footnotesize I} gas 
(total H{\footnotesize I} mass $= 1.6 \times 10^9 M_{\odot}$) 
are instead found along the tidal bridge (Smith et al.~1997). 
A string of young star clusters along the bridge 
is also visible in HST/WFPC2 images (Fig.~2; see also Struck \& Smith 2003), 
although no obvious counterpart is seen in the ULX 
error circle (this is partly because a WFPC2 chip gap 
runs exactly across the circle).
From the level of H$\alpha$ emission in the maps of Smith et al.~(1997) 
and Gonz\'{a}lez-Delgado et al.~(1995), 
the star-formation rate per unit area 
in the western part of the bridge (where it joins 
the stellar ring, near the ULX location) seems to be $\sim 3$--$3.5$ 
orders of magnitude lower than in the nuclear starburst region 
(total H$\alpha$ luminosity from the bridge $\approx 2 \times 10^{39}$ 
erg cm$^{-2}$: Smith \& Struck 2001).
Assuming that the X-ray luminosity scales with the H$\alpha$ emission 
and the star-formation rate (Ranalli et al.~2003; Buat et al.~2002) 
the extended X-ray emission component from the bridge 
would be $\la 10^{38}$ erg cm$^{-2}$, at least 
a factor of 10 fainter than our {\it XMM-Newton}/EPIC 
detection limit. Despite the comparatively weaker star-forming 
activity, the bright ULX is located there rather 
than in the nuclear region. So, the effect of tidal interactions 
on ULX formation must be more complex than a simple increase 
of the star-formation rate.

We can clearly rule out the possibility that the ULX 
be a massive runaway binary formed in the nucleus and then 
migrated to the outer ring. If, as suggested by its luminosity, 
the total mass of the binary system is $> 25 M_{\odot}$, 
we expect kick velocities $< 10$ km s$^{-1}$ (Zezas \& Fabbiano 2002). 
At a distance of $\approx 4$ kpc from the nucleus, 
it would have taken the ULX $> 4 \times 10^8$ yr to get there, 
more than the lifetime of any massive donor stars 
and possibly more than the age of the nuclear starburst itself 
(Lan\c{c}on et al.~2001).

Despite the high column density of atomic hydrogen 
($\ga 10^{21}$ cm$^{-2}$: Smith et al.~1997), 
CO (1--0) observations failed to detect the tidal bridge
(Smith \& Struck 2001). This can be interpreted as 
evidence for the low metal abundance of the gas.
The galactic nucleus is also known to be metal-poor ($\sim 0.2$--$0.4 Z_{\odot}$: 
Garc\'{i}a-Vargas et al.~1997). 
Other bright ULXs have been found in interacting galaxy systems, 
located at or near large, high-density H{\footnotesize I} or H$_2$ complexes, 
but with only a weak level of local star formation. 
For example, the ULX M81 X-9 is associated (Wang 2002) 
with a molecular cloud or ``protogalaxy'' 
near Holmberg IX, with a gas mass $\sim 10^7 M_{\odot}$ 
and star-formation rate $\la 10^{-4} M_{\odot}$~yr$^{-1}$
(Brouillet et al.~1992; Henkel et al.~1993).
In fact, the last close encounter between M\,81 and M\,82 
seems to have left a number of tidal tails and debris (Yun et al.~1994), 
stripped from the metal-poor outer part of M\,81, which are now 
evolving into tidal dwarf galaxies, 
with metal abundances $\sim 0.1$--$0.4 Z_{\odot}$ 
(Boyce et al.~2001; Makarova et al.~2002).

As pointed out by Pakull \& Mirioni (2002), stellar evolution 
in metal-poor environments may favor the formation of ULXs, 
because the mass-loss rate in the radiatively-driven 
wind from the massive O-star progenitor is much reduced 
($\dot{M}_{\rm w} \sim Z^{0.85}$: Vink et al.~2001; 
see also Bouret et al.~2003).
This leads to a more massive stellar core, which may then collapse 
into a more massive BH.
Hence, it may explain the formation of BHs with masses up to 
$\approx 50 M_{\odot}$ 
and isotropic luminosities $\approx 10^{40}$ erg s$^{-1}$.
Tidal dwarfs, bridges and tails stripped from the outer disks 
of colliding galaxies may provide an environment 
for the formation of molecular clouds and stars out 
of metal-poor gas.
However, normal stellar evolution processes alone would 
be unable to explain accreting BHs with isotropic luminosities 
$\approx 10^{41}$ erg s$^{-1}$ such as the one seen 
in NGC\,7714. 
 
BHs with masses of $\sim 10^2$--$10^3 M_{\odot}$ 
may be formed in the core of young (age $\la 3 \times 10^6$ yr), 
massive star clusters, due to the Spitzer instability, runaway 
core collapse and merger of the O stars (Portegies Zwart \& McMillan 2002; 
Rasio et al.~2003).
Deeper optical observations of the ULX field in this galaxy, 
and a more precise determination of its position\footnote{The system 
has never been observed by Chandra.}, 
are needed to investigate the possible association 
with massive young clusters.

Alternatively, BH remnants in this mass range are thought 
to be left over by primordial-metallicity (Population III) stars 
(Madau \& Rees 2001). Most of these BHs could reside in galactic halos: 
assuming halo masses of $6 \times 10^{10} M_{\odot}$ 
and $2 \times 10^{10} M_{\odot}$ for NGC\,7714/15 respectively
(Struck \& Smith 2003), and using the results of Islam et al.~(2003), 
one can estimate the presence of $\sim 200$ such 
intermediate-mass BHs in this system. In an undisturbed galactic halo, 
accretion from the low-density interstellar medium would 
make them too faint to be detected. Could galactic collisions
and mergers create the conditions for some of these BHs 
to become bright (up to their Eddington luminosity $\sim 10^{41}$ 
erg s$^{-1}$), by providing a fuel supply? 
This could in principle occur either via tidal capture 
of a donor star, or via accretion from a dense molecular cloud. 
The tidal capture timescale can be comparable 
with the dynamical timescale of the NGC\,7714/15 interaction only 
for stellar densities $\ga 10^5$ stars pc$^{-3}$ (Fabian et al.~1975; 
Zezas \& Fabbiano 2002), much larger 
than in the collisional stellar ring. 
Hence, this process should be ruled out.

On the other hand, let us consider Bondi-Hoyle accretion from 
an intergalactic molecular cloud. Assuming an efficiency 
$\eta = 0.1$, an isotropic luminosity $\approx 10^{41}$ 
erg s$^{-1}$ requires an accretion rate $\dot{M} \approx 10^{21}$ 
g s$^{-1}$~$\approx 2 \times 10^{-5} M_{\odot}$ yr$^{-1}$.
The Bondi-Hoyle accretion rate (Bondi \& Hoyle 1944; Mirabel et al.~1991) is 
\begin{displaymath}
\dot{M} \approx 2.4 \times 10^{11} (M/M_{\odot})^2\,(n/{\rm cm}^3)\,
(v/10 {\rm ~km~s}^{-1})^{-3}\ \ {\rm g~s}^{-1},
\end{displaymath}
where $M$ is the mass of the accreting BH, 
$n$ is the number density of the molecular gas at large distances from the BH, 
and $v$ is the velocity of the compact object relative 
to the cloud. Choosing $M = 10^{3} M_{\odot}$ and $v = 10$ km s$^{-1}$, 
H$_{2}$ number densities $\approx 2 \times 10^4$ cm$^{-3}$ are required 
to explain the observed luminosity. 
Such densities are typical of molecular cloud cores 
(eg, Plume et al.~1997), with sizes $\sim 0.1$ pc. 
The total molecular gas mass in the NGC\,7714 bridge 
is likely to be $\sim 10^8$--$10^9 M_{\odot}$, depending 
on the metal abundance (Smith et al.~1987; Smith \& Struck 2001). 
Hence, the filling factor of dense molecular cloud cores 
near the western end of the tidal bridge is only $\sim 10^{-4}$.
The probability that a primordial BH could encounter 
one such cloud appears to be low.
Addressing these issues in more details is beyond the scope 
of this work.

\subsection{Nuclear activity}

Our {\it XMM-Newton} study shows that the X-ray  
spectrum of the starburst nucleus contains 
a variable power-law component in addition to non-variable 
thermal plasma emission (Table 2). This suggests that 
there is probably a single accreting point-source 
in the nuclear region contributing a significant fraction 
of the power-law emission, 
at least $3 \times 10^{40}$ erg s$^{-1}$. It had been speculated 
in the past that NGC\,7714 might contain a low-luminosity AGN, 
but it was then shown (O'Halloran et al.~2000) that a normal starburst 
can account for the IR/optical/UV properties.
However, these new observations suggest 
the presence either of an AGN, or of another ULX.
Other cases of galaxies with variable AGN activity 
undetected from their optical colors or spectra 
are discussed in Davis \& Mushotzky (2004) and references 
therein.
A more detailed study of the nuclear starburst, 
combining the new X-ray data with the multiwavelength results 
of Lan\c{c}on et al.~(2001), and Gonz\'{a}lez-Delgado 
et al.~(1995, 1999), is left to further work.

Finally, we note that the NGC\,7714/15 system is one 
of the examples of alignments between a galaxy and two quasars 
discussed by H. Arp and collaborators (Stocke \& Arp 1978). 
Within that scenario, it has been suggested (Arp et al.~2004)
that ULXs and quasars might be related to the same physical 
process of ejection from galactic nuclei.
We point out for completeness that the ULX in NGC\,7714 is also aligned 
with the nucleus and the two quasars. However, we have no elements 
to speculate that this represents more than a coincidence.

\section{Conclusions}

We have used {\it XMM-Newton} to study the interacting galaxy 
system NGC\,7714/15. We have reported here the main properties 
of the two brightest sources: the starburst nucleus and 
an off-nuclear ULX. 
The X-ray spectrum of the off-nuclear source suggests 
that it is an accreting BH, and rules out the possibility 
that it is due to thermal-plasma emission from a hot spot, 
as previously speculated. Its X-ray flux varies by a factor of 2 over 
a six months' interval; the source appears softer 
in the low state, unlike the typical behaviour of Galactic BH candidates 
but in agreement with the behaviour of many ULXs.
Its spectrum in the low/soft state can be fitted by a disk-blackbody 
model with $kT_{\rm in} \approx 1$ keV: this is inconsistent 
with a Shakura-Sunyaev disk, but can be explained 
with a slim-disk model. 

In the high state, its emitted isotropic luminosity is $\approx 6 \times 
10^{40}$ erg s$^{-1}$ in the $0.3$--$12$ keV band, 
implying a bolometric luminosity $\approx 1.5 \times 10^{41}$ erg s$^{-1}$ 
from a reasonable extrapolation of the power-law spectrum 
(photon index $\Gamma \approx 2$). BH masses of $\sim$ 
a few $10^2$--$10^3 M_{\odot}$ would be required to satisfy the Eddington 
limit. Furthermore, variability on timescales of $\approx 2$ hr is 
detected in the high state.

The ULX is located at the junction of the tidal bridge 
(consisting of gas and young stars) with the collisional outer 
ring (consisting of an old stellar population, with no gas). 
We have pointed out that ULXs are often found 
in tidally interacting systems, associated 
with metal-poor molecular clouds, tidal dwarfs, 
or H{\footnotesize I} structures formed 
in the galactic collision.

The nucleus has an X-ray luminosity $\approx 10^{41}$ erg s$^{-1}$ 
in the $0.3$--$12$ keV band. Thermal plasma emission 
contributes for $\approx 3 \times 10^{40}$ erg s$^{-1}$, 
constant over the two observations, 
and is probably extended (marginally resolved 
in the EPIC/MOS images). A point-like power-law component 
contributes for $\approx 5 \times 10^{40}$ erg s$^{-1}$ 
and $\approx 8 \times 10^{40}$ erg s$^{-1}$ in the two observations.
The power-law component in the X-ray spectra of starburst nuclei 
is generally due to unresolved high-mass X-ray binaries. 
The amount of variability in our case implies that one single 
source contributes for at least $3 \times 10^{40}$ erg s$^{-1}$.
This suggests that there is either a hidden AGN or another ULX 
in the nuclear region.

\begin{acknowledgements}
We thank Manfred Pakull for the
preparation of the observations and for discussions.
\end{acknowledgements}

\end{document}